\documentclass[11pt]{article}
\usepackage{latexsym}  % for Symbol fonts
\usepackage{amssymb}
\usepackage{graphicx}
\usepackage{amsmath}

\begin{document}
%\hfill  2020.02.20
\vspace{5mm}

%%%%%%%%%%%%%%%%%%%%%%%%%%%%%%%

\begin{center}
%{\Large\bf  Detection of extensive cosmic ray air showers by measuring radio emission }
{\large\bf  An electron linac with high beam intensity over 1-Ampere for disposal of radioactive waste } 

\vspace{10mm}
{\bf Y. Kawashima $^{a}$ \footnote{Corresponding author \\ Email address: kawa@rcnp.osaka-u.or.jp  or bnlkawa@gmail.com (Y. Kawashima)} , 
T. Asaka $^{b}$,  H. Ego $^{c}$ and M. Hara $^{d}$ }

 {\it $^{a}$Research Center for Nuclear Physics (RCNP), 10-1 Mihogaoka, Ibaraki, Osaka, 567-0047, JAPAN \\
 $^{b}$SPring-8, 1-1-1 Kouto, Sayo-cho, Sayo-gun, Hyogo, 679-5198, JAPAN\\
 $^{c}$KEK, 1-1 Oho,Tsukuba, Ibaraki 305-0801, JAPAN\\
 $^{d}$RIKEN, 2-1 Hirosawa, Wako, Saitama 351-0198, JAPAN\\}
\end{center}

\begin{quotation}
In order to dissipate long-lived radioactive waste, not only high flux proton accelerator but also electron linac have been proposed. A proton accelerator directly induces nuclear fission and mutation. On the other hand, electron beam has two processes: production of gamma rays through bremsstrahlung and produced gamma rays are available for the $(\gamma, n)$ reaction. Protons dissipate nuclei and simultaneously newly produce radioisotopes of by-products. Since an electron linac produces less radioactive wastes, we focus only an electron linac with high beam intensity more than 1-Ampere. To reduce the yield of by-products for radioisotopes as less as possible, we accelerate electron beam with the energy less than 30 MeV. The accelerated beam intensity is designed to be more than 1-Ampere. To realize an electron linac with high intensity, the operation for the linac is not pulsed one, but continuous wave. To accelerate electron beam, we install higher-order-modes (HOMs) free normal conducting cavities to suppress beam instabilities.
\end{quotation}

\vspace{5mm}
%%% Section 1.0
%%   \eqno(1.1)
{\bf I. \ Introduction }
%{\large\bf I. \ Introduction }
% \section{Introduction}
\vspace{2mm}

In these days, the radioactive waste of fission products is one of the biggest problems for human beings. In order to solve the problem, scientists have to deeply consider how to dissipate long-lived radioactive waste. To reduce long-lived radioactive waste, there is a facility to research using proton beam \cite{view1}. On the other hand, there is another proposal to transmute radioisotopes (RI) with long lifetime by using photons \cite{Matsumoto}. One can see that a feasibility study for the calculation of $( \gamma, n)$ reaction was carried out \cite{Matsumoto}. The paper \cite{Matsumoto} mentions that high flux $\gamma-rays$ over $10^{19} /s $ significantly incinerate long-lived isotopes such as $^{90}Sr$ and $^{135}Cs$.  There is a unique proposal, which combines photons produced by an electron linac and reactor \cite{Brown}.

 The proton beam is a useful tool to dissipate nuclear wastes. However, the proton beam easily produces other radioisotopes as by-products through the strong interaction process. Therefore, we do not discuss the proton beam anymore.
 To transmute long-lived radioactive waste, we select the method of an electron linear accelerator, which has been proposed by the reference \cite{Matsumoto}. However, the electron linac with high beam intensity more than 1-Ampere so far has not been easy method. Nowadays, the B-factory at KEK stably stores electrons and positrons more than 1-Ampere \cite{view2}. A new electron linac with the beam intensity more than 1-Ampere should be based on the technology developed for the B-factory. It seems to be a common thing that the operation for a linear accelerator is pulsed one. In order to increase the beam intensity more than 1-Ampere, we should adopt a continuous wave (CW) operation. 

Since the binding energies for nuclei are mainly distributed near the center of 8 MeV, it seems to be good enough to set the electron beam energy at around 10 MeV, and also from the point of suppressing the yield of radioisotopes as by-products \cite{Brown}. However, we increase the beam energy up to from 20 to 30 MeV, because the cross section for the Bremsstrahlung drops steeply near the maximum electron beam energy as shown in Fig.1 \cite{Petrukhin}. Based on the calculation result as shown in Fig.1, we can evaluate the total photon number with more than 8 MeV and per $ 1/(g \cdot cm^2) $ to be 
around $ 10^{17} $ to $10^{18} $ photons at the beam energy of 20 MeV and 30 MeV and the current of 1-Ampere. Thus we set the maximum beam energy at 30 MeV. In actual operation, electron beam energy is variable from 20 MeV to 30 MeV. 

% Figure 1 should be inserted here
%  Fig_1
\begin{figure}[ht]
\begin{center}
  \includegraphics[scale=0.4]{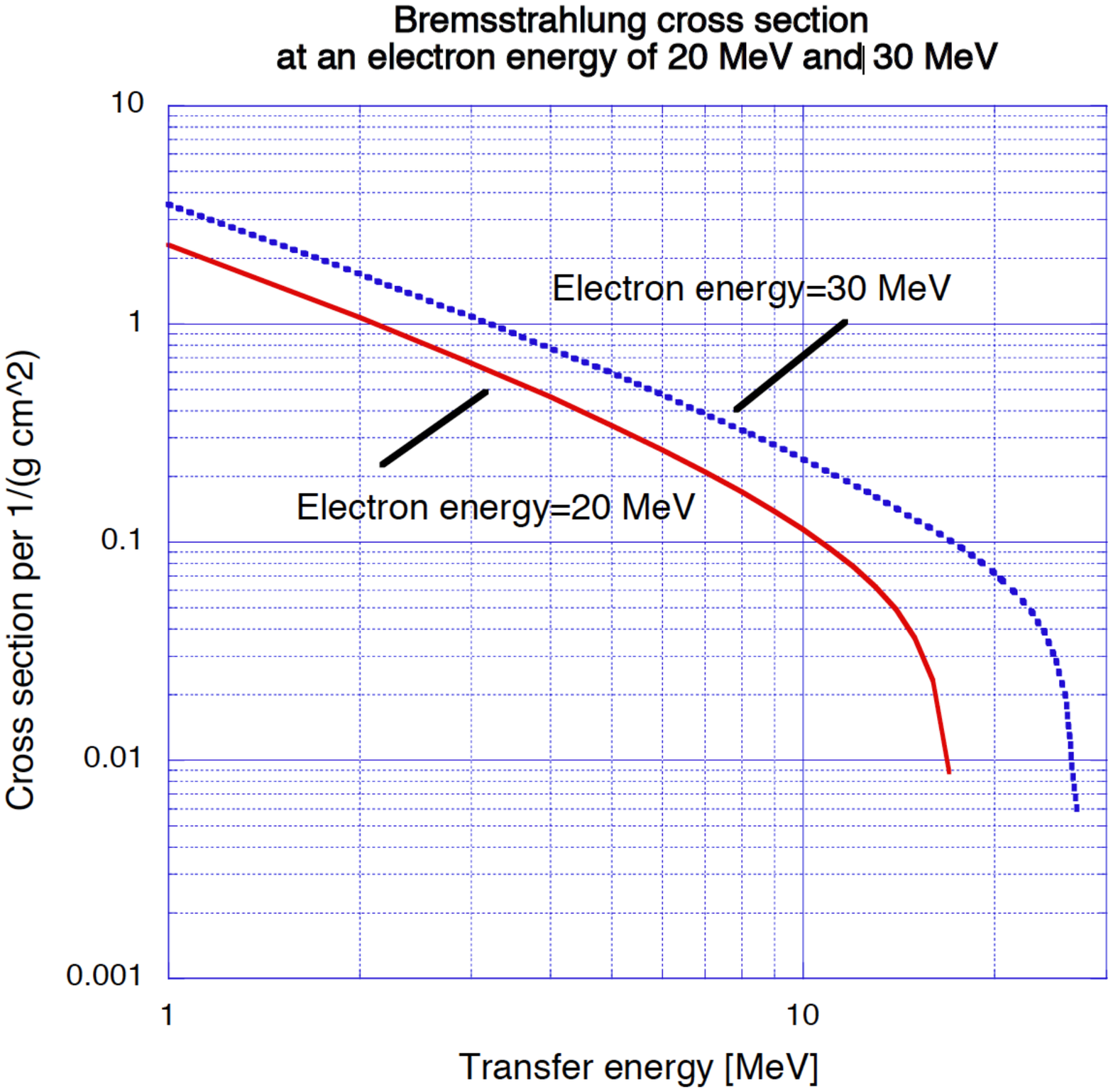}
  \caption{Cross sections per $1/(g \cdot cm^2) $ for the Bremsstrahlung at the electron energy of 20 MeV and 30 MeV.  }
  \label{fig1}
  \end{center}
\end{figure}

To make an electron linac, we investigate which cavity is more useful for a high flux linac: a super conducting or a normal conducting cavity as  higher-order-modes (HOMs) free cavity. In the case of a super conducting cavity, the generated voltage becomes very high. Therefore, beam loading problem occurs for the super conducting cavity. As an example, total radio-frequency (RF) power $ P $ is expressed by beam current $ I $ and generated voltage $ V $,
% equation (1)
\begin{equation}
\ P= I \times V \
\end{equation}
The maximum RF power is limited by an amplifier, in general, klystron output power. When the voltage generated by a cavity is set higher, beam current is limited. To increase the beam current more than 1-Ampere, the cavity voltage should be set lower.

 In general, a single superconducting cavity generates more than 1 MV even at around a few watts input RF power, because the shunt impedance is very high. To accelerate 1-Ampere electron beam by a single superconducting cavity, the required RF power is calculated  $1 [MV] \times 1 [A] = 1 [MW] $. However, we do not have an input coupler, that withstands up to 1 MW input RF power. As long as we know, the maximum available input coupler was developed by F. Naito et al. \cite{Naito} at KEK and it withstands over 850 kW as an input RF power. Thus, we do not discuss about superconducting cavities anymore.
 A normal conducting cavity with lower voltage is much better to accelerate high current electrons. In order to accelerate an intense electron beam, higher-order-modes (HOMs) free cavity should be selected to suppress beam instability.
 Available klystrons for an electron linac are generally designed for pulsed operation and the maximum output RF power attains up to around 80 MW. However, such klystrons are not useful to accelerate higher flux beam. We need to choose continuous wave operable klystrons as an RF amplifier. In this paper, we describe the design for high intensity electron linac more than 1-Ampere.
 
%% Section 2
\vspace{5mm}
{\bf II. \ Normal conducting cavities with HOMs free}\\
%{\large\bf II. \ Normal conducting cavities with HOMs free } \\
%\vspace{2mm}
% \section{Normal conducting cavities with HOM free}
%% subsection 2.1
% \subsection{\bf{In case of ARES cavity}}
\indent
{\bf (1) In case of ARES cavity}

Available HOMs free and normal conducting cavity was designed and developed at the KEK B-factory. It was named ARES cavity \cite{Yamazaki}. By using the ARES cavity, it is possible to accelerate electron beam more than 1-Ampere. We list the specification of ARES cavity in Table 1. It should be noticed that the Q-value is designed to be higher to suppress Robinson instability and is useful for a storage ring.

% Table 1
\begin{table}[ht]
\begin{center}
\begin{footnotesize}
%\begin{tiny}
\caption{The specification for ARES cavity.}
% \begin{tabular}{llr}
 \begin{tabular}{lcc}
% \begin{tabular}{|c| l| l| l| l| l| l| l| l|}
\hline
\hline
 Unloaded Q-value  &  1.18E5    \\
 \hline
    $ R_{SH}$  (shunt impedance) &  1.75E6 ohms  \\
  \hline
    \hline
 \end{tabular}
\end{footnotesize}
%\end{tiny}
\end{center}
\end{table}

As for a high power and continuous wave (CW) klystron, a candidate is an E3732 produced by CANON \cite{view3}. The maximum RF power is 1.2 MW and the available frequency is 508.58 MHz. Combining the ARES cavity and the klystron, we can design an electron accelerator as shown in Fig. 2. Since available input coupler is acceptable for 850 kW as the maximum input power [6], we need to divide RF power from a klystron as shown in Fig.2.

% Figure 2 should be inserted here
%  Fig_2
\begin{figure}[ht]
\begin{center}
  \includegraphics[scale=0.4]{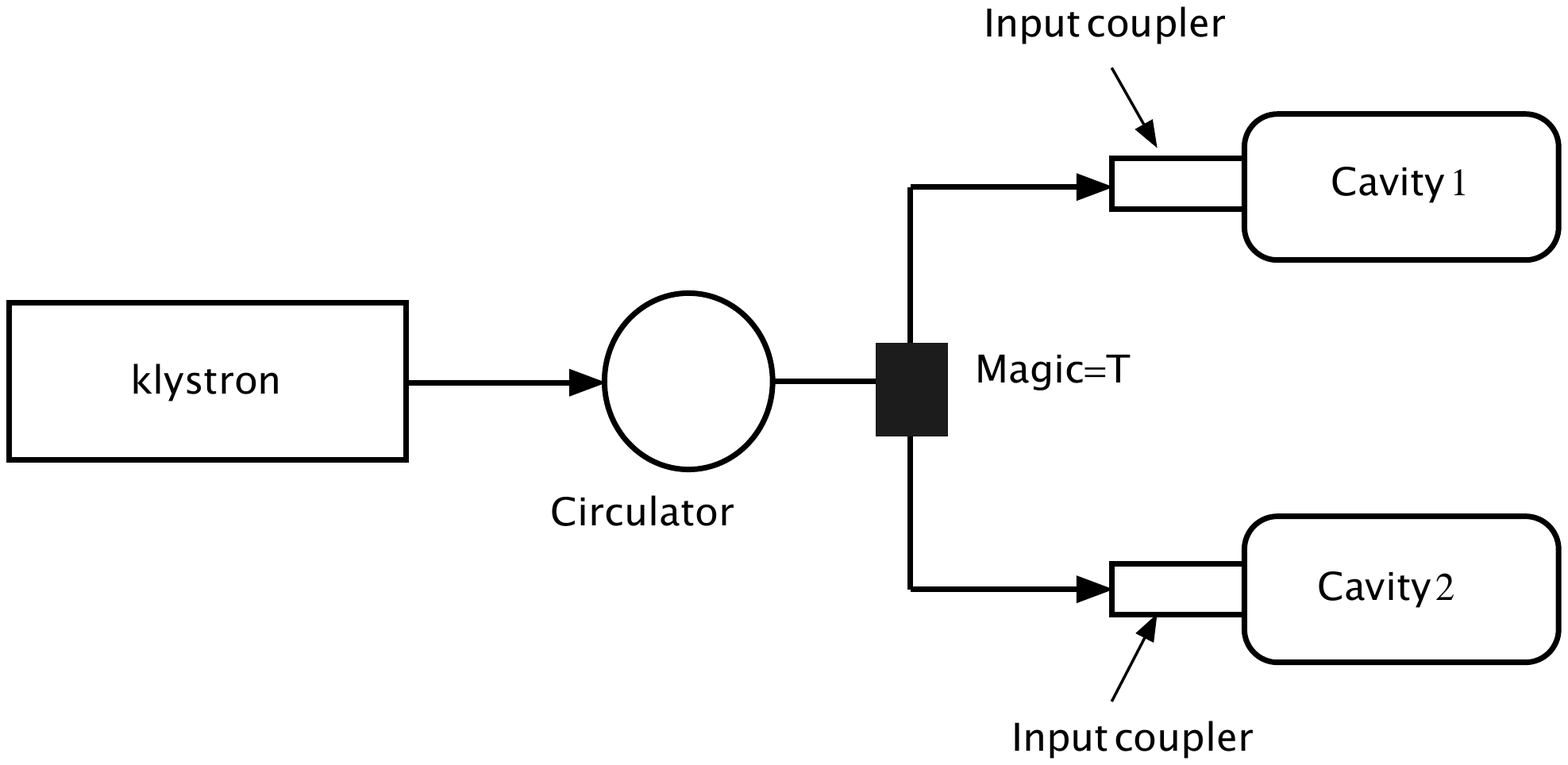}
  \caption{Layout of a klystron and two cavities for an actual operation. }
  \label{fig2}
  \end{center}
\end{figure}

When ARES cavity generates 0.5 MV, the maximum RF power as a beam loading in the layout of Fig. 2 becomes 1 MW to accelerate 1-Ampere electron beam. To decrease the maximum RF power below 1 MW, we set the operation voltage for ARES cavity at 0.4 MV. In order to generate 0.4 MV, required RF power P is calculated by using Table 1,
% equation (2)
\begin{equation}
P=\frac{\left | V^{2} \right |}{R_{SH}}=\frac{(0.4\times 10^{6})^{2}}{1.75\times 10^{6}}=91.4 \ [kW].
\end{equation}
And we also calculate the beam loading per one cavity,
% equation (3)
\begin{equation}
P_{L}=0.4[MV]\times 1[A]=400 \ [kW].
\end{equation}
Thus, total RF power as shown in Fig.2 is obtained by
% equation (4)
\begin{equation}
Total \hspace{0.03in}RF\hspace{0.03in} power=(91.4\hspace{0.03in}kW+400\hspace{0.03in}kW)\times 2=982\hspace{0.03in} \ [kW]
\end{equation}
The continuous operation for a klystron operating nearly maximum RF power is not common thing. Because the lifetime for a klystron becomes shorter and shorter. Taking into account the lifetime for a klystron, we can set the initial generated voltage at lower level of 0.3 MV. However, to accelerate electron beam up to 30 MeV, the number of RF components increases. As a result, total cost becomes much higher. Therefore, we consider another solution in the next section. 

\vspace{2mm}
%% subsection 2.2
%\subsection{\bf In case of EGO cavity }
{\bf (2) In case of EGO cavity}
%\vspace{2mm}

A new HOMs free cavity was designed and developed. We named EGO cavity \cite{Ego1}. We list the specification for the EGO cavity in Table 2. 

% Table 2
\begin{table}[ht]
\begin{center}
\begin{footnotesize}
%\begin{tiny}
\caption{The specification for EGO cavity.}
%\begin{tabular}{llr}
\begin{tabular}{lcc}
%\begin{tabular}{|c| l| l| l| l| l| l| l| l|}
 \hline
\hline
 Unloaded Q-value  &  6.0E4    \\
  \hline
    $ R_{SH}$  (shunt impedance) & 6.8E6 ohms  \\
  \hline
   \hline
\end{tabular}
\end{footnotesize}
%\end{tiny}
\end{center}
\end{table}

EGO cavity has higher shunt impedance comparing Table 2 with Table 1. To generate 0.4 MV, required RF power is calculated by
% equation (5)
\begin{equation}
P=\frac{(0.4 \times 10^{6})^{2}}{6.8 \times 10^{6}}=23 \hspace{0.03in}[kW].
\end{equation}
The RF power calculated in (5) becomes smaller than the value obtained in (2). Beam loading is also calculated by
% equation (6)
\begin{equation}
Beam \hspace{0.03in} loading=1 \hspace{0.03in} [A] \times 0.4 \hspace{0.03in} [MV]=400 \hspace{0.03in} [kW].
\end{equation}
We finally obtain RF power per an EGO cavity, 423 kW. In actual operation, we also consider the layout of Fig.2 and obtain total RF power,
% equation (7)
\begin{equation}
Total \hspace{0.03in} RF \hspace{0.03in} power=(400 \hspace{0.03in} [kW] \times 23 \hspace{0.03in} [kW]) \times 2= 846 \hspace{0.03in} [kW].
\end{equation}
We compare two cavities: ARES and EGO. We might conclude that the EGO cavity is much better to accelerate electron beam more than 1-Ampere.
 We summarize RF power consumption for ARES and EGO cavities in Fig.3. 

% Figure 3 should be inserted here
%  Fig_3
\begin{figure}[ht]
\begin{center}
  \includegraphics[scale=0.4]{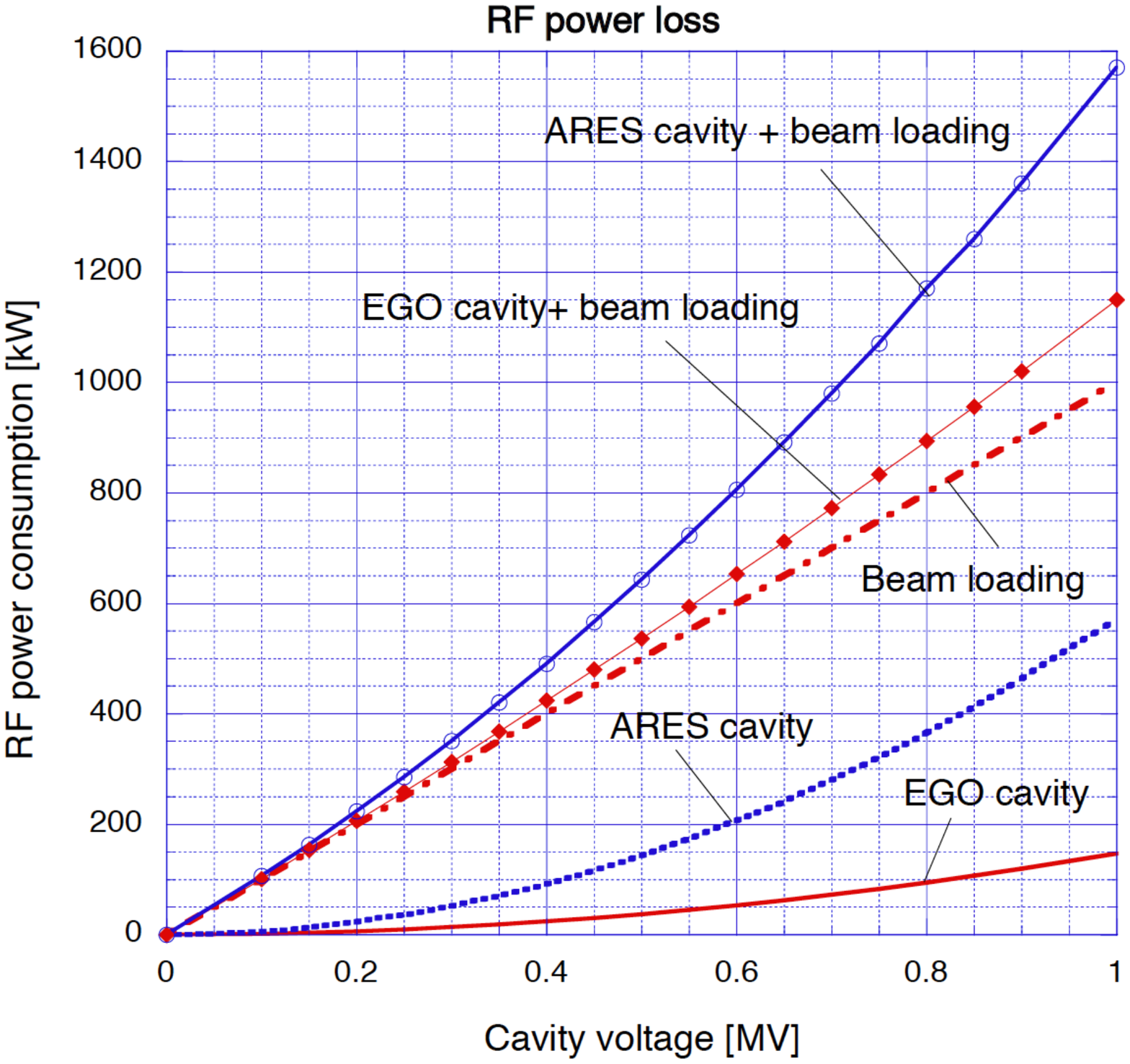}
  \caption{Summary for RF power consumption of ARES and EGO cavities.}
  \label{fig3}
  \end{center}
\end{figure}

%% Subection 2.3
\vspace{2mm}
%% subsection 2.3
%\subsection{\bf Total RF components to accelerate electron beam over 1-Ampere and 30 MeV }
{\bf (3) Total RF components to accelerate electron beam over 1-Ampere and 30 MeV} 
%\vspace{2mm}
% \subsection{ \bf Total RF components to accelerate electron beam over 1-Ampere and 30 MeV}

We have mentioned to accelerate electron beam up to 30 MeV. In order to complete the purpose, we need to list up total RF components. The main parts have been shown in Fig.2, where electron beam obtains the energy of 0.8 MeV. To accelerate electron beam up to 30 MeV, we need 38-set as shown in Fig.2. Let us summarize the total RF components for ARES and EGO cavities. We list up all RF components in Table 3.

% Table 3
\begin{table}[ht]
\begin{center}
\begin{footnotesize}
%\begin{tiny}
\caption{Total numbers for all RF components.}
% \begin{tabular}{llr}
%\begin{tabular}{lcc}
 \begin{tabular}{|c| l| l| l| l| l| l| l| l|}
\hline
\hline
 Name of parts  &  Numbers of items    \\
 \hline
   \hline
     Klystoron  & 38   \\
  \hline
       High  power  equipment  for  klystron &  19   \\
        \hline
            Circulator  & 38   \\
            \hline
               Input  coupler  or  ceramic  window  & 76  \\
                \hline
                   Waveguide  &  including Magic-T   \\
                    \hline
                       Cavity & 76  \\
                         \hline
                          \hline
\end{tabular}
\end{footnotesize}
%\end{tiny}
\end{center}
\end{table}

%% Section 3
\vspace{5mm}
{\bf III. \ Electron injector }
%{\large\bf III. \ Electron injector }
%\vspace{2mm}
% \section{Normal conducting cavities with HOM free}
% \section{Electron injector}

In order to supply more than 1-Ampere electron beam, we have designed electron injector system by using PARMELA simulation code \cite{Young}. The beam operation is not pulsed one, but continuous wave (CW). One can see total layout of injector section as shown in Fig.4. The conventional thermionic-gun stage for electron generation is charged at -90 kV. The total length for the injector section is almost 6 meters long. We set up 8 cavities, which are operated at 508.58 MHz CW. Cavities 1 and 2 work to bunch a coasting beam. In order to focus electron beam,  nine magnetic lenses are placed. These magnetic fields are distributed around 350 Gauss. The generated electron beam from the thermionic-gun at the voltage of -90 kV is accelerated up to 1.6 MeV after the cavity 8. The results obtained by the PARMELA simulation are summarized in Table 4. Since generated voltages from cavity1 to 4 are lower, we do not need high power klystron. Instead of a klystron, inductive output tubes (IOT) are useful to supply lower RF power at the frequency of 508.58 MHz. On the other hand, a klystron for cavity 5 to 8 is available to supply RF power. 

% Figure 4 should be inserted here
%  Fig_4
\begin{figure}[ht]
\begin{center}
  \includegraphics[scale=0.4]{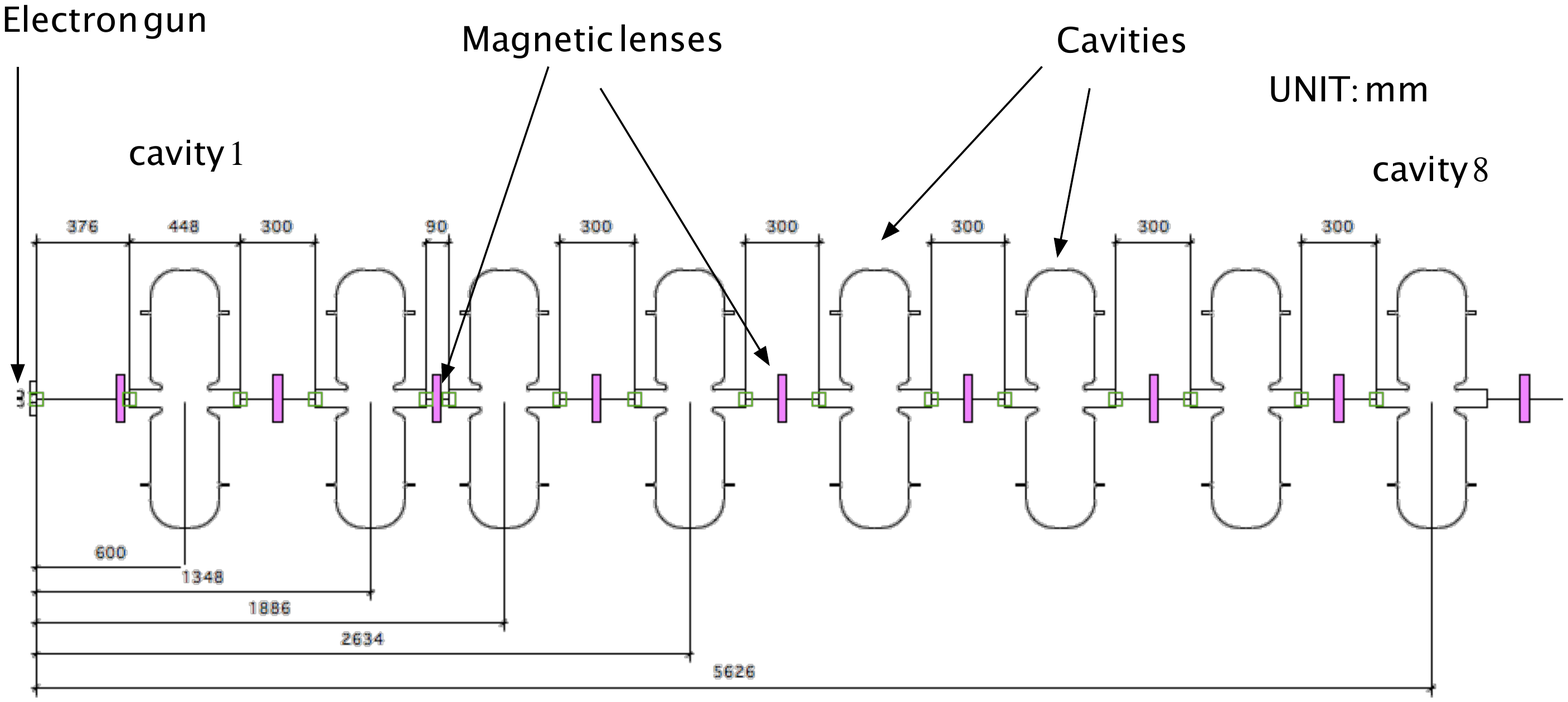}
  \caption{The schematic layout for electron gun system designed by PARMELA.}
  \label{fig4}
  \end{center}
\end{figure}

  %% Section 4
\vspace{5mm}
{\bf IV. \ Evaluation for electric power consumption }\\
%{\large\bf IV. \ Evaluation for electric power consumption \\}
%\vspace{2mm}
% \section{Normal conducting cavities with HOM free}
%\section{Evaluation for electric power consumption} 
%% subsection 4.1
\indent
{\bf (1) Electron injector section}
%\subsection{\bf Electron injector section}

In order to evaluate the electric power consumption, we assume the ever-developed single-cell cavity \cite{Ego2}. The shunt impedance is about 5 Mohms. We can calculate RF power consumption for each cavity. 
% equation (8)
\begin{equation}
Cavity \ 1 = 92 [W]. 
\end{equation}
% equation (9)
\begin{equation}
Cavity \ 2 \ to \ 4= 24.5 \ [kW],  \ Total \ RF \ power=73.5 \  [kW]. 
\end{equation}
% equation (10)
\begin{equation}
Cavity \ 5 \ to  \ 8= 32  \ [kW], \ Total \ RF \ power=128 \ [kW]. 
\end{equation}
Total RF power consumption is summed up and obtains 202 kW. We assume the conversion efficiency 50 \% from electric power to RF power. The required electric power for the electron injector section is evaluated 404 kW. 

\vspace{2mm}
%% subsection 4.2
{\bf (2) Main linac section}
%\subsection{\bf Main linac section}

Total number of klystrons are 38 as shown in Table 3. To accelerate 1-Ampere electron beam, two cavities require 846 kW obtained in the equation (7). Thus, we can get the total RF power consumpton using both values. 
% equation (11)
\begin{equation}
Total \hspace{0.03in} RF \hspace{0.03in} power= 846 \hspace{0.03in} [kW] \times 38 \hspace{0.03in} [kW] = 32 \hspace{0.03in} [MW].
\end{equation}
We need to make an assumption for the conversion efficiency 50 \%, too. We obtain 64 MW as a total electric power. Even though we combine electric powers for both sections: electron injector and main linac sections, the electric power consumption in the electron injector section is negligible small comparing to the main linac section.
% Table 4
\begin{table}[ht]
\begin{center}
\begin{footnotesize}
%\begin{tiny}
\caption{Total numbers for all RF componets.}
%\begin{tabular}{llr}
%\begin{tabular}{lcc}
\begin{tabular}{|c| l| l| l| l| l| l| l| l|}
\hline
 \hline
 Name of parts  &  Obtained simulation results    \\
 \hline
   \hline
     Beam Transmission efficiency  & $ \sim 50 \% $  \\
  \hline
       Bunch length &  $< 300ps $  at output  \\
        \hline
            Emittance  & $ < 200 \pi \cdot mm \cdot mrad $ at output   \\
            \hline
               Energy  & 1.6 MeV  \\
                \hline
                   Cavity 1 & 21.5 kV, phase=96 degrees  \\
                    \hline
                       Cavity 2 & 350 kV, phase=-5 degrees  \\
                        \hline
                          Cavity 3 & 350 kV, phase=-5 degrees  \\ 
                           \hline
                               Cavity 4 & 350 kV, phase=100 degrees  \\    
                             \hline            
                                   Cavity $5 \sim 8 $ & 400 kV, phase= 0 degree \\ 
                                   \hline
                                    After \ cavity 8  &  2.45 nC/280 ps (= 8.8 A at peak ) \\
                            \hline
                             \hline
\end{tabular}
\end{footnotesize}
%\end{tiny}
\end{center}
\end{table}

%% Section 5
\vspace{5mm}
{\bf  V. \ Summary}
%\section{Summary}

We have designed an electron linac to dissipate long-lived radioactive waste. The designed linac with the beam current of 1-Ampere generates the photons of $10^{18}/s$ with the energy of more than 8 MeV per $ 1/(g \cdot cm^2)$. In order to obtain the photon flux more than $10^{18}/s$, it is easy to increase photon conversion thickness or target thickness from  $ 1/(g \cdot cm^2) $ to  $ 10/(g \cdot cm^2) $.

 If funding problem is solved, this linac is easily made by using current technology developed at mainly KEK. As for the electric power consumption, the required total electric power must be less than 100 MW including cooling water system as well as quadrupole magnets to focus electron beam. 
 When one sets the beam energy at lower values of 20 MeV and also suppresses a klystron output power less than 850 kW, beam intensity is attainable up to 1.5-Ampere. Thus we are able to change beam current as well as beam energy.
 When long-lived radioactive waste is irradiated by the electron beam with the maximum energy of 30 MeV, the target is heated up by the electron beam with the maximum power of 30 MW. Therefore the target cooling system becomes very serious. We need to develop the target cooling system in future.

\vspace{5mm}
\section*{Acknowledgements}
% \begin{Acknowledgements}
We are indebted to Professor K. Nakai for his continuous support and M. Fujiwara for giving several references.  We would like to express our thanks to Y. Yamazaki and T. Kageyama for their contributions to develop HOM free cavity.

\end{document}